\documentclass[onecolumn]{article}    
\usepackage{graphicx} %
\usepackage{amsmath} 
\usepackage{amssymb}  
\usepackage{lipsum}
\usepackage{xcolor}
\usepackage{color}
\usepackage{cite}
\usepackage[normalem]{ulem}
\usepackage{graphicx}          
\DeclareMathOperator{\E}{E}
\DeclareMathOperator{\trace}{tr}

\DeclareMathOperator{\vect}{vec}
\DeclareMathOperator{\rank}{rank}
\DeclareMathOperator{\diag}{diag}

\renewcommand{\theequation}{\thesection.\arabic{equation}}

\begin{document}
\title{Stable Reduced-Rank VAR Identification}                                                

\author{Xinhui Rong and Victor Solo
\footnote{Authors are with School of Electrical Eng. $\&$ Telecommunications, 
UNSW, Sydney, Australia. }  
}    
  
\maketitle

\begin{abstract}                          
The vector autoregression (VAR) has been widely used in system identification, econometrics, natural science, and many other areas. However{{,}} when the state dimension becomes large the parameter dimension explodes. So rank reduced modelling is attractive and is well developed. But a fundamental requirement in almost all applications is stability of the fitted model. And this has not been addressed in the rank reduced case. Here, we develop, for the first time, a closed-form formula for an estimator of a rank reduced transition matrix which is guaranteed to be stable. We show that our estimator is consistent and asymptotically statistically efficient and illustrate it in comparative simulations.
\end{abstract}

\section*{Copyright Statement}
© 2024 Xinhui Rong and Victor Solo. This work has been accepted to Automatica for publication under a Creative Commons Licence CC-BY-NC-ND. 

\section{Introduction}
\setcounter{equation}{0}
\label{intro}
The vector autoregression (VAR) is 
widely applied in control, signal processing, time series and econometrics. 
The lag-one VAR (VAR(1)) is more general than it seems
since any higher order VAR can be written as a state space model
whose state equation is a VAR(1) \cite{Lutke05}\cite{VanO96}. 
 {{We refer the readers to Section 2 for technical details.} }
More recently, the VAR(1) has been used in 
reinforcement learning \cite{Sutt18}\cite{Bert19}\cite{Matn19}\cite{Rech19}.

High dimensional time series have now become very common
across many disciplines, e.g. 
epidemiology \cite{Weik16}, financial economics \cite{Rein98}.
Rank reduced modelling has gained increasing attention 
\cite{Rein98}\cite{RCHN20}\cite{RCHN21}, 
 {{
since in many cases the system dynamics can be well 
captured in a reduced-rank model 
where some (if not most) poles are near zero 
so are playing almost no part in the dynamics. 
Another important advantage of reduced-rank modelling 
is that in the full-rank modeling, 
higher-dimensional vector data lead to an explosion in 
the number of parameters, which often exceeds the 
number of available vector data, while 
the reduced rank factorization of the system matrix 
reduces the number of parameters significantly
(See Section 4 for details). 
Sparsity is also a natural approach to deal with systems with huge dimensions. 
However, it is not obvious how to guarantee stability with sparsity. 
}}

In physical system, stability is crucial.
And this has led to a continuing development
of stability guaranteed estimators, e.g. for
state space modelling.
A common approach employs a {\it perturbation minimization} procedure, 
where a preliminary unstable least-squares estimator is stabilized 
by the smallest additive perturbation. 
For example, 
Mari et al. \cite{Mari00} consider a semidefinite programming ({{SDP}}) problem 
with a Lyapunov stability constraint. 
The problem can be solved by standard linear matrix inequality (LMI) methods. 
However, the computation becomes prohibitive with higher-dimensional states.
Miller and de Callafon \cite{Mill13} extend the {{SDP}} method to 
include more eigenvalue constraints, 
besides stability. 
Boots et al. \cite{Boot07} and the first algorithm in \cite{Mall08} 
use singular value constraints instead of eigenvalue constraints, 
which can be too conservative. 
The second algorithm in \cite{Mall08} uses a line search method 
based on the gradient sampling and 
the reliability is sensitive to the user-defined step size parameter. 
Tanaka and Katayama \cite{Tana05} and Jongeneel et al. \cite{Jong23} 
project the unstable estimator onto a stable region 
by solving linear quadratic regulator (LQR) problems. 
Jongeneel et al. \cite{Jong23} show that 
the computational efficiency is greatly improved in higher-dimension applications, 
and provide error analysis and statistical guarantees. 
 {{Orbandexivry et al. \cite{Orba13} use a barrier projection method. 
Gillis et al. \cite{Gill19} find a new characterization for stable matrices 
and use the gradient search. 
Noferini and Poloni \cite{Nofe21} convert the perturbation minimization problem 
to an equivalent form by Schur factorization and use the Riemannian gradient search. }}

Other approaches include,
Chui and Maciejowski \cite{Chui96}
who iteratively augment the unstable estimator 
until its largest eigenvalues have the modulus of the user-defined parameter. 
However, this method distorts the estimator and introduces additional bias. 
Van Gestel et al. \cite{VanG01} introduce a regularization term 
to the least squares problem 
so that the variance of the estimators is reduced and in the meantime, 
the stability is guaranteed. 
However, the regularization is too conservative. 
Umenberger et al. \cite{Umen18} consider the state space modelling 
and take the maximum likelihood approach to guarantee stability. 
However, their method is computationally expensive. 

In recent work,
we \cite{Rong23} 
solved a state space estimation problem by
using a Burg-type {\it forwards-backwards (FB) optimization}
\cite{Burg75}\cite{Stra77}\cite{Nutt77} on the error residuals and 
remarkably find  a closed-form stable solution that is computationally cheap
and involves no tuning parameters.

Among the above works, 
{{\cite{Orba13}\cite{Gill19}\cite{Nofe21} treat the `nearest stable matrix' problem 
without system context, }}
\cite{Jong23} %
deal with pure VAR(1), and 
the others consider the state space setup 
where subspace identification methods are needed before fitting a VAR(1) model \cite{VanO96}. 
However, none of the above stability-enforced methods 
deals with the reduced rank modeling.

In this paper, we 
extend the work in \cite{Rong23},
using the FB approach
to obtain a reduced rank (RR)-VAR(1) with guaranteed stability. 
The FB approach generates an estimator that 
is no more computationally expensive than
the least squares RR method of \cite{Rein98}.
Further, no tuning parameters are required.

The rest of the paper is organized as follows. 
We first introduce the forwards and backwards VAR(1) models 
in Section 2, 
and then develop for the first time:
\begin{enumerate}
\item[(i)] in Section 3, a closed-form stable estimator for the full-rank VAR(1), 
\item[(ii)]  {{i}}n Section 4, based on (i), a closed-form stable estimator for the RR-VAR(1), 
\item[(iii)] in Section 5, statistical consistency, and asymptotic efficiency
for the new estimator.
\end{enumerate}
The results are illustrated in comparative simulations in Section 6. 
Section 7 contains conclusions.

We use the following notations. 
 {$\rho(A)$ is the spectral radius of $A$. 
$\Vert A\Vert=\sqrt{\trace(AA')}$ is the Frobenius norm of $A$. 
$A>(\geq)0$ means $A$ is positive (semi-)definite. 
 {{`$\xrightarrow{p}$' means convergence in probability.} }
w.p.1 means with probability 1.}

\section{VAR(1) and the Backwards Model}
\setcounter{equation}{0}
In this section, we review the (forwards) VAR(1) and its associated backwards model.
The VAR(1) is a Markov process generated by
\begin{align}
\label{f-var}
y_t = F y_{t-1} + w_t,\quad t = 1,\dotsm,T,
\end{align}
where $y_t$ is the observed $n-$vector time series, 
$w_t$ is a zero-mean driving white noise 
with a non-singular covariance matrix $Q$, 
and $F$ is the transition matrix, assumed to be stable. 
The stability of $F$ 
ensures the existence of a steady state variance matrix $\Pi$ 
which obeys a discrete-time (DT) Lyapunov equation
 {$
\Pi = F\Pi F' + Q.
$}
A sufficient condition for $\Pi$ to be positive definite is that $Q$ is positive definite.
We introduce the steady state lag-one cross-covariance 
$\Pi_{10} = \E[y_ty_{t-1}']=F\Pi=\Pi_{01}'$, 
so that
$F=\Pi_{10}\Pi^{-1}$.
We further introduce the associated correlation matrix
$R_F=\Pi^{-\frac12} \Pi_{10}\Pi^{-\frac12}=\Pi^{-\frac12} F\Pi^{\frac12}$
which has the same eigenvalues as $F$.

Associated with the VAR(1) is a stationary
backwards model \cite{KAIL00} 
\begin{align*}
y_{t-1} = F_by_t+w_{b,t-1}, \quad t = T, \dotsm, 1, 
\end{align*}
where $F_b$ is the backwards transition matrix
and $w_{b,t-1}$ is a white noise with covariance matrix $Q_b$
which is statistically independent of $y_t$.
We note that \cite{KAIL00}
\begin{align*}
F_b &= \E[y_{t-1}y_t']\E^{-1}[y_ty_t] = \Pi_{10}'\Pi^{-1} = \Pi F' \Pi^{-1}\\
{\mbox{and }} Q_b &= \E[(y_{t-1}-F_by_t)(y_{t-1}-F_by_t)'] \nonumber\\
	&= \Pi - F_b\Pi_{10} - \Pi_{10}'F_b' + F_b\Pi F_b'\nonumber\\
	&= \Pi - \Pi_{01}\Pi^{-1}\Pi_{10} \nonumber\\
	&= \Pi - F_b\Pi F_b'.
\end{align*}
An elementary  argument shows that 
$F_b$ has the same eigenvalues of $F$. 
Thus  $F_b$ is stable iff $F$ is stable. 
The following results are used below.

{\bf Theorem 1. Converse Lyapunov Result.} 
If $\Pi$ is positive definite and $Q$ is positive semi-definite then
any eigenvalue $\lambda$ of $F$ has $|\lambda|\leq 1$. If $Q$ is positive definite
then $|\lambda|<1$.

{\it Proof}.
{Let $v$ be left eigenvector of $F$ with eigenvalue $\lambda$. Then
denoting $(v^{\ast})'$ as $v^H$ we find 
 {$v^H\Pi v=|\lambda|^2v^H\Pi v+v^H Qv
\Rightarrow	(1-|\lambda|^2)v^H \Pi v=v^H Qv\geq 0\Rightarrow |\lambda|\leq 1.$} 
Clearly if $Q$ is positive definite we get $|\lambda|<1$.
\hfill$\square$

{\bf Theorem 2}. $Q_b$ has full rank iff $Q$ has full rank.

{\it Proof}.
First note that $ Q_b=\Pi Q_c\Pi$ where $ Q_c=\Pi^{-1}-F'\Pi^{-1} F$.
Then clearly $ Q_b$ has full rank iff $ Q_c$ has full rank.
Suppose for some vector $v\neq 0$, we have $ Q_b v=0$.
Then $ Q_c\Pi v=0$. Thus, $\Pi^{-1} \Pi v=F'\Pi^{-1} F\Pi v$,
i.e. $v=F'\Pi^{-1} F\Pi v$. Note that we cannot have $F\Pi v=0$ since then
$v=0$ which is a contradiction.

Now consider that
\begin{align*}
\begin{array}{rrcl}
&Q\Pi^{-1}&=&I-F\Pi F'\Pi^{-1}\\
\Rightarrow
&Q\Pi^{-1} F\Pi&=&F\Pi- F\Pi F'\Pi^{-1} F\Pi\\
\Rightarrow
&Q\Pi^{-1} F\Pi v&=&F\Pi v-F\Pi F'\Pi^{-1} F\Pi v\\
&&=&
F\Pi v-F\Pi v=0 {{,}}
\end{array}
\end{align*}
which is a contradiction and proves `if'; then `only if' follows by running the argument
in reverse.
\hfill$\square$

\section{Stable Estimators for full-rank VAR(1)}  
\label{full}
\setcounter{equation}{0}
Here we review earlier work of \cite{Nutt77}\cite{Stra77}\cite{Rong23} 
and develop a new result 
on the full-rank case.
Given data $y_t ,t=0,\cdots, T$ set
$Y_0 = [\begin{matrix}y_0&\dotsm&y_{T-1}\end{matrix}]$ and
$Y_1 = [\begin{matrix}y_1&\dotsm&y_T\end{matrix}],$
and define the sample covariances
$S_{ij} = \frac1TY_iY_j', \quad i,j\in\{0,1\}$.
We also introduce the forwards and backwards residual mean squared errors
\begin{align*}
S_{w,f}(F) 	&= \frac1T(Y_1 - FY_0)(Y_1 - FY_0)'\\
		&= S_{11} - FS_{01} - S_{10}F' + FS_{00}F'\\
S_{w,b}(F) 	&= \frac1T(Y_0 - F_bY_1)(Y_0 - F_bY_1)'\\
		&= S_{00} - F_bS_{10} - S_{01}F_b' + F_bS_{11}F_b',
\end{align*}
where
$F_b=PF'P^{-1}$, where $P$ is 
a consistent estimator of $\Pi$ to be chosen. 
Here $F$ is no longer the true value. 
The least squares estimator 
$\hat F_{LS}  =  S_{10} S_{00}^{-1}$ minimises $\trace( S_{w,f})$,
but it is NOT guaranteed to be stable. 

Continuing, \cite{Nutt77}\cite{Stra77} considered the minimizer of the sum
of weighted forwards and backwards sample mean squared errors
\begin{align*}
J(F;P)	=\trace\{P^{-1}(S_{w,f}+S_{w,b})\}.
\end{align*}
This yields the following remarkable result.

{\bf Theorem 3A}. \cite{Stra77,Rong23,Nutt77}. 
 {{For any positive definite $P$, }}the minimizer of $J(F;P)$ is $\hat F$,
 {{which}} obeys the Sylvester equation below 
and is stable
\begin{align*}
\hat F S_{00}P^{-1} + S_{11}P^{-1}\hat F = 2S_{10}P^{-1}.
\end{align*}

{\it Remarks.}
\begin{enumerate}
\item[(a)] This result is not explicitly stated in \cite{Stra77} but is rather a special case
of his results. Further the argument in \cite{Stra77} is
very hard to follow. Partly for that reason,  {in \cite{Rong23}
we} gave a simple direct proof. Note that Theorem 4B below
has nearly the same proof.
\item[(b)] The result only holds with the weighting matrix $P$
used to define $F_b$, not
for other weighting matrices.
\item[(c)] Using vec algebra we can get a closed-form solution.
\end{enumerate}
We now have the further remarkable closed-form result.

{\bf Theorem 3B}. 
Set $P= S_{11}$. Then the stable estimator 
\begin{align*}
\hat F=\hat F_{11} =2S_{10}(S_{00}+S_{11})^{-1} {{.}}
\end{align*}
{\it Proof}. Follows from Theorem 3A by setting $P= S_{11}$.

{\it Remarks}.
\begin{enumerate}
\item[(d)] Below we find that the choice $P= S_{11}$
enables our new results.
\item[(e)] Note that
consistency and asymptotic efficiency
 {{have}} been established for $\hat F_{LS} $ in \cite{Lutke05}.
One can then show the same for $\hat F$.
We omit 
details, concentrating here on
the reduced rank case.
\end{enumerate}

\section{Stable Reduced Rank VAR(1) Estimation}
\setcounter{equation}{0}
We now consider the case where $F$ has rank $m<n$.
Then we can factor $F=A_{n\times m} B_{m\times n}$. 
We then have the following result.

 {{\bf Theorem 4A. 
 Reduced Rank Least Squares.}}\cite[Section 2.3]{Rein98}.
 {{The reduced rank least squares } }
 {{$
\min_{\rank(F)=m}J_{LS}:J_{LS}=\trace( S_{11}^{-1} S_{w,f}),
$} }
has solution
$\hat F_{RLS} =  S_{11}^{\frac12}\hat V_{*,m}\hat V_{*,m}' S_{11}^{-\frac12}\hat F_{LS}$
where $\hat V_{*,m}=[\begin{matrix}\hat v_{*,1}&\dotsm&\hat v_{*,m}\end{matrix}]$
and $\hat v_{\ast,r}${{'s}} are the `top' $m$ eigenvectors of
$\hat R_* =  S_{11}^{-\frac12} S_{10} S_{00}^{-1} S_{01} S_{11}^{-\frac12}.$
The noise covariance estimator is
$\hat Q_{RLS}= S_{11}^{\frac12} (I-\hat V_{*,m}\hat D_{*,m}^2\hat V_{*,m}') S_{11}^{\frac12}$
where $\hat D_{*,m}^2=\diag(\hat \lambda_{*,k})$ contains the top $m$ eigenvalues of $\hat R_*$.

{\it Remarks}.
\begin{enumerate}
\item[(f)] $\hat F_{RLS}$ is not guaranteed to be stable.
\item[(g)] We note for future reference that $\hat R_*$ has the same
eigenvalues as $\hat L_*= S_{10} S_{00}^{-1} S_{01} S_{11}^{-1}$.
\item[(h)] We can re-express the results in terms of the SVD of 
$\hat G_*= S_{11}^{-\frac12}  S_{10} S_{00}^{-\frac12}=\hat V_*\hat D_*\hat U_*'$ 
since $\hat R_*=\hat G_*\hat G_*'$.
It follows from the Biename-Cauchy-Schwarz inequality that
the singular values of $\hat G_*$ are  {{less than $1$,}} w.p.1. 
 {{Thus, $\hat Q_{RLS}$ is positive definite w.p.1.}}
\item[(i)] In \cite{Rein98}, a different formula for $\hat Q_{RLS}$ is given 
but it is straightforward to show it equals the one given here.
\end{enumerate}
We would now like to set up a FB version of this problem.
But it turns out that for general $P$ there is no simple solution.
Fortunately, if we choose $P= S_{11}$ we can find a simple guaranteed stable estimator.

{\bf Theorem 4B.  {{Stable Reduced Rank Estimator.}}} 
\label{F-rr1}
The solution to
$\min_{\rank(F)=m}J(F; S_{11})$ is given by
$\hat F_R =  S_{11}^{\frac12} \hat V_m\hat V_m'  S_{11}^{- {{\frac12}}}\hat F_{11}$ where
$\hat V_m=[\begin{matrix}\hat v_{1}& \dotsm& \hat v_m\end{matrix}]$
and $\hat{v}_r${{'s}} are the `top' $m$ eigenvectors 
(corresponding to the top $m$ eigenvalues in $\hat D_m=\diag(\hat\lambda_k)$) of
 {$
\hat R = 2S_{11}^{-\frac12}S_{10}(S_{00}+S_{11})^{-1}S_{01}S_{11}^{-\frac12}. 
$}
Further $\hat F_R$ is stable.  

{\it Proof}. See the appendix. 


\section{Asymptotic Analysis}
\setcounter{equation}{0}
\label{asym}
Here we show the consistency and central limit theorem (CLT)
for $\hat F_R $ by reducing them to results for $\hat F_{RLS}$.
For asymptotic results, we need stronger assumptions.
There are a wide range of possibilities, but
to keep things simple, while still retaining
reasonable generality we use:

{\bf Assumption A1}.
$w_t$ in (2.1) are iid with finite fourth moments,
and $Q$ is positive definite.

{\bf Assumption A2}. $F$ has rank $m$ and the 
non zero eigenvalues of $F$ are distinct.

This enables the following result 
for $\hat F_{RLS}$ (where $F$ is again the true value).

{\bf Theorem 5A}. Assuming $m=\rank(F)$ is known : \\ 
(i) Under A1: $ S_{00}\xrightarrow{p} \Pi$ and $ S_{11}\xrightarrow{p} \Pi$.\\
(ii) Under A1: $\hat F_{LS}\xrightarrow{p} F \equiv  S_{10}\xrightarrow{p} F\Pi$.\\
	Under A1, A2: $ R_F =\Pi^{-\frac12} F\Pi^{\frac12}$ has rank $m$ and:
\begin{enumerate}
\item[(iii)] $\hat V_{*,m}\xrightarrow{p} V_m$ the eigenvectors of $ R_F  R_F '$.\\
$\hat U_{*,m}\xrightarrow{p} U_m$ the eigenvectors of $ R_F ' R_F $.\\
$\hat\lambda_k\xrightarrow{p} \lambda_k\mbox{ and }\hat\lambda_{*,k}\xrightarrow{p}\lambda_k$ the $k^{th}$ largest eigenvalue of $ R_F  R_F '$.
\item[(iv)] $\hat G_*\xrightarrow{p}  R_F =\Pi^{-\frac12} F\Pi^{\frac12}$. 
\item[(v)] $\hat F_{RLS}\xrightarrow{p} F=AB$. 
\item[(vi)] $\sqrt{T} \vect(\hat F_{RLS}'-F')\Rightarrow Z\sim N(0,\Sigma)$ where \\
$\Sigma=MWM'$ with $M=[(I\otimes B'),(A\otimes I)]$, 
 {{$A=Q^{\frac12}V_m, B=V_m'Q^{-\frac12}F$, }}
and $W$
(which is complicated) is given in  {\cite[(2.36)]{Rein98}}. 
\end{enumerate}
{\it Proof}. (i),(ii) can be found in \cite[Chapter 11]{Ham94}.
(v),(vi) are from  {\cite[Theorem 2.4 and Section 5.2]{Rein98}}.
For (iii) first note that from (i),(ii) $\hat L_*\xrightarrow{p} L=F\Pi F'\Pi^{-1}$
which has the same eigenvalues as $R= R_F  R_F '$. The second part follows
in the same way.
Next,
since distinct eigenvalues and eigenvectors are continuous functions 
of the underlying matrix entries
\cite{Neud99} then{{\footnote{
If $\xi_T\xrightarrow{p} \xi$ and $f$ is continuous then $f(\xi_T)\xrightarrow{p} f(\xi)${{.}}}
$\hat V_{*,m}\xrightarrow{p} V_m$}}, 
containing the `top'  {{eigenvectors}} of $R$.
Further, the eigenvalues  {{converge in probability}} as stated.
(iv) now follows from (iii). \hfill$\square$

To continue for $\hat F_{R}$  {{asymptotics}}, we need some lemmas.

{\bf Lemma 5B}. 
Let $\Gamma_t=\E[y_ty_t']$. Then
$\Vert \Gamma_t-\Pi\Vert_2\leq c[ {{\rho(F)}}]^{2t}$ for some constant $c$.

{\it Proof}. Taking variances in (\ref{f-var}) gives
$\Gamma_t=F \Gamma_{t-1} F'+Q$. Subtracting the $\Pi$ equation from this gives
$\Gamma_t-\Pi=F(\Gamma_{t-1}-\Pi)F'$.
Iterating this gives $\Gamma_t-\Pi=F^t D (F')^t$ where $D=\Gamma_0-\Pi$.
Now recall Gelfand's theorem \cite{HORN13}: $\Vert F^t\Vert^{1/t}\rightarrow  {{\rho(F)}}$,
so that $\lim\sup_t \Vert (\Gamma_t-\Pi)/ {{[\rho(F)]}}^{2t}\Vert^{1/t}\leq c$.
\hfill$\square$

The result follows from this.

{\bf Lemma 5C}. 
Under A1: $\sqrt{T}\Vert S_{11}- S_{00}\Vert\xrightarrow{p}  0$.

{\it Proof}. We have 
\begin{align*}
\begin{array}{rrcl}
 &S_{11}- S_{00}&=&\frac1{T}[y_Ty_T'-y_0y_0']\\
\Rightarrow
&\Vert S_{11}- S_{00}\Vert^2&\leq&\frac{2}{T}(\Vert y_T\Vert^2+\Vert y_0\Vert^2)\\
\Rightarrow &\sqrt T \E\Vert S_{11}- S_{00}\Vert^2
&\leq &\frac2{\sqrt T}[\trace(\Gamma_T)+\trace(\Gamma_1)] {{.}}
\end{array}
\end{align*}
The second term  {{converges to $0$}} while the first term  {{converges to $0$}} from Lemma 5B. 
The result then follows.
\hfill$\square$

{\bf Lemma 5D}.  Under A1, A2: 
$\sqrt{T}\Vert\hat F_{LS}-\hat F_{11}\Vert\xrightarrow{p} 0$.

{\it Proof}. We have
\begin{align*}
\begin{array}{rcl}
&&\sqrt T(\hat F_{LS}-\hat F_{11})\\
&=&
\sqrt T( S_{10} S_{00}^{-1}-2 S_{10}( S_{00}+ S_{11})^{-1})\\
&=&
\sqrt T 2 S_{10}[(2 S_{00})^{-1}-( S_{11}+ S_{00})^{-1}]\\
&=&
\sqrt T  S_{10} S_{00}^{-1}[ S_{00}+ S_{11}-2 S_{00}]( S_{00}+ S_{11})^{-1}\\
&=&
\hat F_{LS}[\sqrt T( S_{11}- S_{00})]( S_{11}+ S_{00})^{-1} {{.}}
\end{array}
\end{align*}
Now from Theorem 5A and Lemma 5C 
each term converges in probability, so the product also converges.
But $\sqrt T( S_{11}- S_{00})\xrightarrow{p} 0$ yielding the result.
\hfill$\square$

{\bf Theorem 5E}. 
Under A1, A2,\\
(i) $\hat F_R \xrightarrow{p} F$.\\
(ii) $\sqrt T(\hat F_{RLS}-\hat F_R )\xrightarrow{p} 0$.\\
(iii) $\sqrt T\vect(F_R '-F')\Rightarrow Z\sim N(0,\Sigma)$.

{\it Proof}. (i) follows from (ii) and Theorem 5A(v).
(iii) follows from (ii) and Theorem 5A(vi).
(ii) is proved in the appendix.
\hfill$\square$

\section{Simulations}
\setcounter{equation}{0}
\label{sim}
We now show simulations to illustrate our stable estimator $\hat F_R$. 
Since there is no existing algorithm that guarantees a stable RR-VAR(1) estimator, 
we compare our estimator 
with the standard, reduced-rank, least square estimator $\hat F_{RLS}$. 
We call our method the FB method and the latter the LS method. 
We expect that the LS method will generate unstable estimates while FB does not, and also that LS and FB will have similar computational complexity given their formulae.

We run two sets of simulations:\\
 {$\bullet$ Low dimensional state $n=6$. \\
$\bullet$ High dimensional state, up to $n>3000$.}\\
The latter study represents increasingly common practical examples.

\subsection{Simulation Design}

Firstly, we specify 
$F=\left[\begin{smallmatrix}F_0&0\\0&0_{3\times3}\end{smallmatrix}\right]\otimes I_{p}$, where 
 {{$I_p$ is the $p$-dimensional identity matrix with}} $p$ to be chosen and 
$F_0 =\left[\begin{smallmatrix}0.99 & -0.1&0\\0.1&0.99&0\\0&0&0.95\end{smallmatrix}\right]$. 
So, there are $3p$ zero eigenvalues and 
$3p$ non-zero eigenvalues repeating at $0.99\pm\jmath0.1$ and  $0.95$, 
very close to the unit circle. 
This is a typical case when a stability-guaranteed algorithm is needed. 
This is a similar setup as in \cite{Jong23}.

Secondly, we let the length $T$ of the time series be a multiple of model order $n$, 
so that the computational complexity 
for both LS and FB is $O(n^3)$ (assuming a standard matrix multiplication method). 
Also, note that in the one-dimensional case, 
the precision $\frac{|\hat F_{LS}-F|}{|F|}$ decays at a rate of $\sqrt{T}$. 
Thus, we consider $T/n=l^2\Rightarrow T=l^2n$. 
The values of $l$ are best determined by system time constants. 
However, here we take a trial-and-error method and let 
$T/n=2^2,6^2,10^2\Rightarrow T=4n,36n,100n$. 

Thirdly, we choose $p=2^k\Rightarrow n=6\times2^k$, 
so the logarithm of the computational time 
is proportional to $k$. 

Throughout this section, we assume the true model order $m$ is known,
 {and }we take the noise covariance $Q=I_n$.

\subsection{Low Dimensional Simulations}
We let $k=0\Rightarrow p=1\Rightarrow n=6$, 
and the rank $m=3$. 
We consider $3$ time series lengths
$T/n=2^2,6^2,10^2\Rightarrow T=24,216,600$. 
We simulate $N=1000$ realizations for each $T$ and get the estimates 
$\hat F_R $ and $\hat F_{RLS}$.

We first plot the estimated pole locations in Fig. \ref{loc}. 
We only plot the first $50$ estimates due to crowdedness. 
It is sufficient to plot the poles with zero or positive imaginary parts. 
Fig. \ref{loc} gives a rough impression of the pole distributions. 
It is observed that as $T$ increases, the poles of the estimates from both algorithms 
cluster closer around the true poles. 
However, the LS estimates have poles outside the unit circle while 
all poles of our FB estimates are inside the unit circle. 

We further plot the histogram of pole magnitudes from the $1000$ repeats in Fig. \ref{polehis}. 
The same conclusions can be drawn as from Fig. \ref{loc}. 
At $T=600$, both histograms have peaks at the true pole magnitudes, 
indicating good performance. 
However, we detect $25.9\%, 6.9\%, 0.2\%$ unstable LS estimates 
for $T=24,216,600$, respectively, 
while our FB method always guarantees stability. 

\begin{figure}[t]
\begin{center}
\includegraphics[width=12cm]{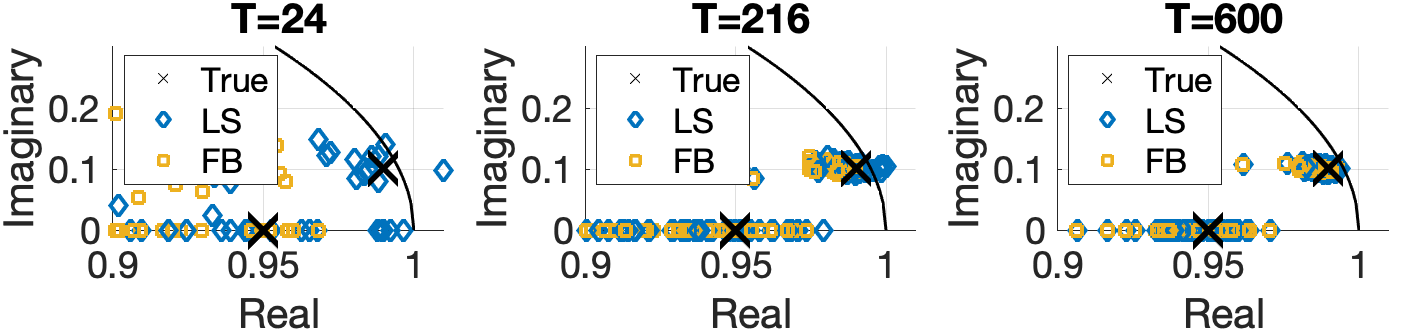}    
\caption{ {Pole locations of the first $50$ repeats. }}  
\label{loc}                                 
\end{center}                                 
\end{figure}

\begin{figure}[t]
\begin{center}
\includegraphics[width=12cm]{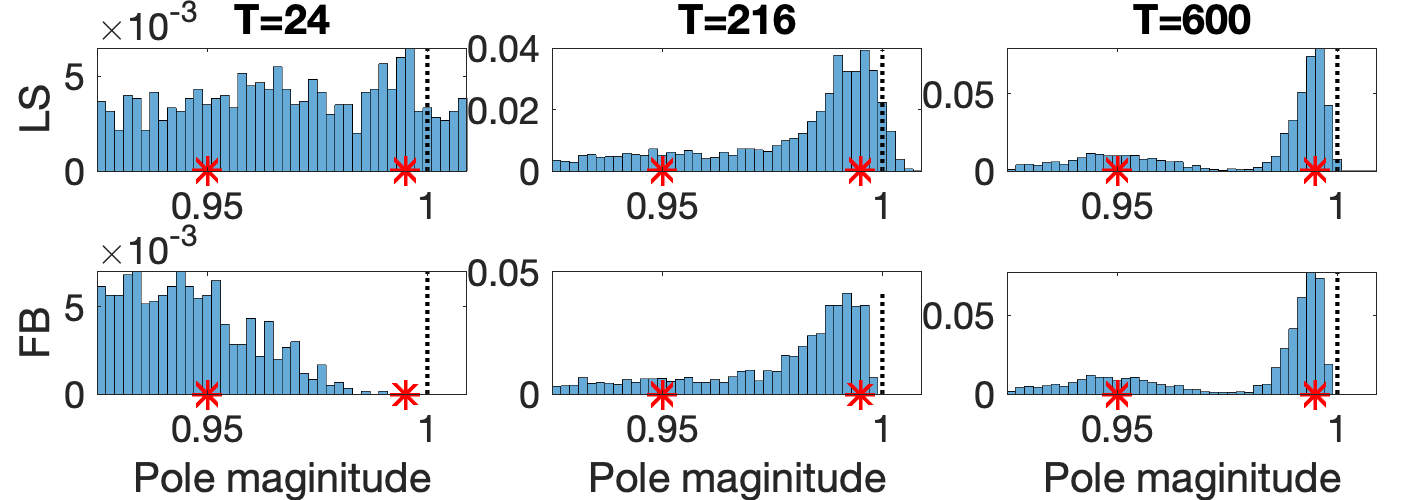}    
\caption{ {Pole magnitude histograms of the $1000$ repeats: $*$ are the true poles. 
Unstable poles for LS for each $T$: $25.9\%, 6.9\%, 0.2\%$.}}  
\label{polehis}                                 
\end{center}                                 
\end{figure}

We introduce the relative estimation error (which are plotted as percentages) 
 {
$ {{e_{RLS}}}=\frac{\Vert \hat F_{RLS}-F\Vert}{\Vert F\Vert}$,  
and prediction error 
$ {{\epsilon_{RLS}}}=\frac{\Vert Y_1-\hat F_{RLS} Y_0\Vert-\Vert Y_1 -\hat F_{LS} Y_0\Vert}
{\Vert Y_1-\hat F_{LS} Y_0\Vert}$}, 
{{and define anagolously $e_{RFB},  \epsilon_{RFB}$ for $\hat F_{R}$. }}
Note that  {{the full-rank}} $\hat F_{LS}$ minimizes the Frobenius prediction error, 
so $\epsilon_{ {{RFB}}},\epsilon_{ {{RLS}}}>0$. 
We {plot the error histograms in Fig. \ref{errlowhis}.}
It is observed that that FB method has competitive performance against LS 
even in small sample cases, whilst guaranteeing stability. 

 {{We have also compared 
 the full-rank estimators $\hat F_{LS}$ and $\hat F_{11}$. 
Space limit precludes additional figures but  
we observe that $\hat F_{LS}$ (resp. $\hat F_{11}$) has worse estimation errors for large $T$ 
compared to the RR estimators 
with the medians $75.7\%, 17.6\%, 10.4\%$ (resp. $79.4\%,17.9\%,10.5\%$), 
for $T=24, 216, 600$, respectively. 
But they have better prediction errors since $\hat F_{LS}$ minimizes such error 
and for $\hat F_{11}$ the medians are $1.7\%, 0.043\%, 0.0047\%$ 
for $T=24,216,600$, respectively. 
We refer the readers to simulations in \cite{Rong23} for comparisons of existing full-rank 
stable estimators. }}

\begin{figure}
\begin{center}
\includegraphics[width=10cm]{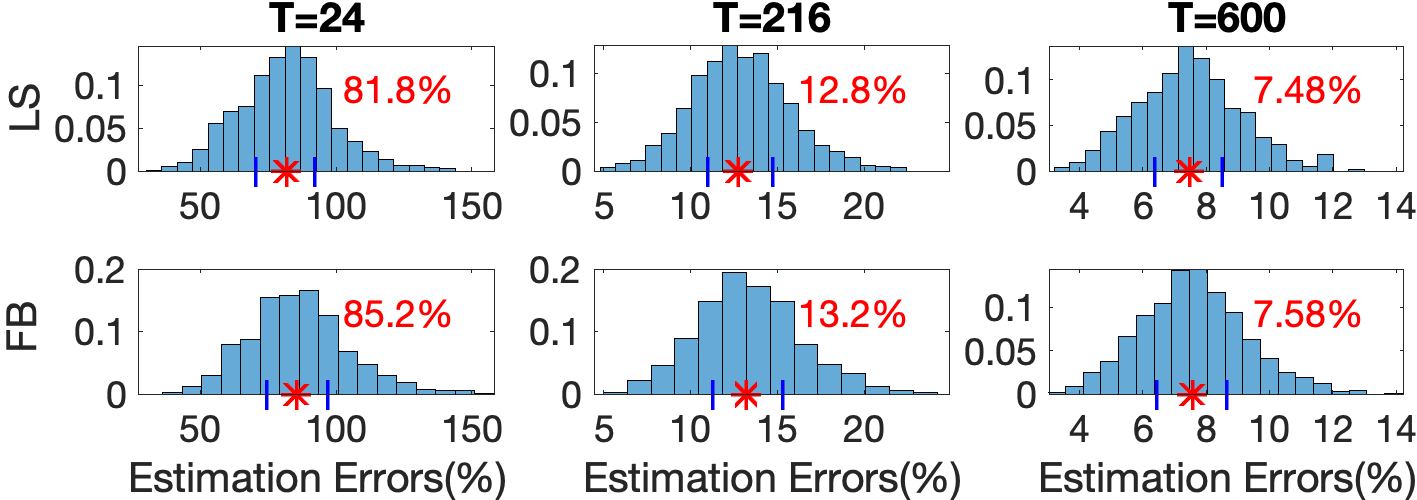}    \\
{(a) Estimation errors.\\
\includegraphics[width=10cm]{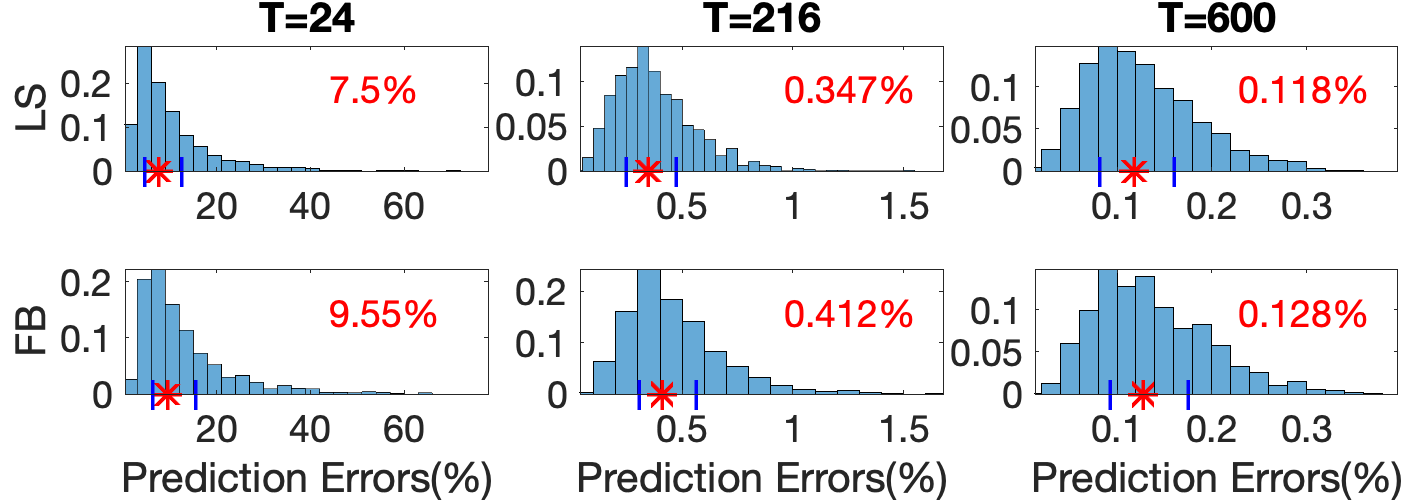}    \\
(b) Prediction errors.
\caption{
Histograms of estimation errors $e_{ {{RLS}}},e_{ {{RFB}}}$ 
and prediction errors $\epsilon_{ {{RLS}}},\epsilon_{ {{RFB}}}$: 
the red `$*$' 
marks and the numbers at the right upper corners 
are the medians and the blue `$|$' marks are the 
upper and lower quantiles.
}\label{errlowhis}                                 
}  
\end{center}                                 
\end{figure}

\subsection{High Dimensional Simulations}
While most stability-guaranteed algorithms are computationally intensive 
(again, these are for full-rank VAR(1)),  
and scale badly to high dimensional data, 
the FB method is not and has the same computational complexity as the LS. 
We demonstrate this feature here. We also compare the estimation and prediction errors.

We consider $k=1,2,\dotsm,9$, so that $n= 12,\dotsm, 3072$. 
We take the time series lengths $T=36n$ and $T=100n$ 
and simulate $N=50$ realizations for each $n$. 
We expect that the computational times $t_{comp}$ for LS and FB be similar 
and that $\log t_{comp}$ be linear with $\log n$. 

We plot the computational times against the model order $n$ in Fig. \ref{compt} in log scale. 
We only plot the case of $T=100n$ 
because the plots of $T=36n$ are very similar. 
It is easily observed that FB is as efficient as LS. 
For the largest model order $n=3072$, the average computational time is only about $49.6$s. 
To the best of our knowledge, 
there exists no (full-rank) stability-guaranteed algorithm 
that can handle such a high model order with achievable computational resources, 
including time and memory.

For the estimation accuracy, 
we compare the estimation errors $e_{ {{RLS}}},e_{ {{RFB}}}$ and the prediction errors 
$\epsilon_{ {{RLS}}},\epsilon_{ {{RFB}}}$ 
 with $T=36n$ and $T=100n$. 
 {{Results are plotted against $k=5,\dotsm,9$ (so the model orders are $n=192,\dotsm,3072$) 
in box plots in Fig. \ref{errFhigh} and Fig. \ref{errprehigh}}}. 
It is interesting to observe that, 
first, LS and FB still have very similar errors in the higher-order cases, 
and second, the error medians converge to constants as $n$ grows 
with $T$ being a multiple of $n$.

\begin{figure}
\begin{center}
\includegraphics[width=7.5cm]{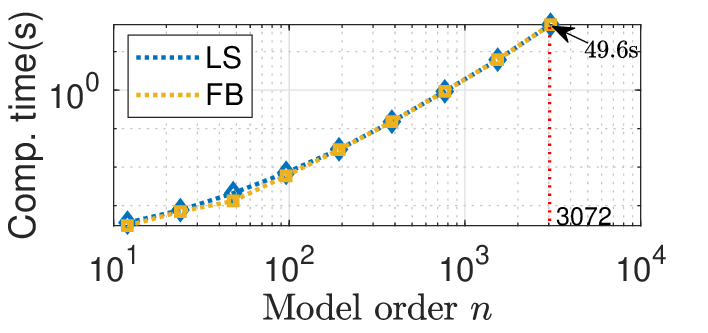}    
\caption{Computational times $t_{comp}$ in log scale: 
$t_{comp}$ for LS and FB are almost identical and $\log(t_{comp})$ is almost linear with $\log n$. 
The average computational time is about $49.6$s for model order $n=3072$.}  
\label{compt}                                 
\end{center}                                 
\end{figure}

\subsection{Simulations Summary}
We conclude that our FB estimator guarantees stability, 
has low computational complexity, and 
has competitive accuracy for both short and long length time series, 
and for both low and high model orders.

\section{Conclusions}
\setcounter{equation}{0}
\label{con}
In this paper, we developed, for the first time, 
a simple, closed-form, stability-guaranteed estimator 
for the reduced-rank VAR(1). It is
based on a forwards-backwards least squares criterion. 
We also gave an asymptotic analysis
showing that the new estimator is consistent and asymptotically efficient. 
Finally, we showed simulations demonstrating 
the competitive accuracy and computational efficiency of the new stable estimator. 

In the future, 
we will develop a
rank selection criterion for medium and large sample sizes. 

\begin{figure}
\begin{center}
\includegraphics[width=7.2cm]{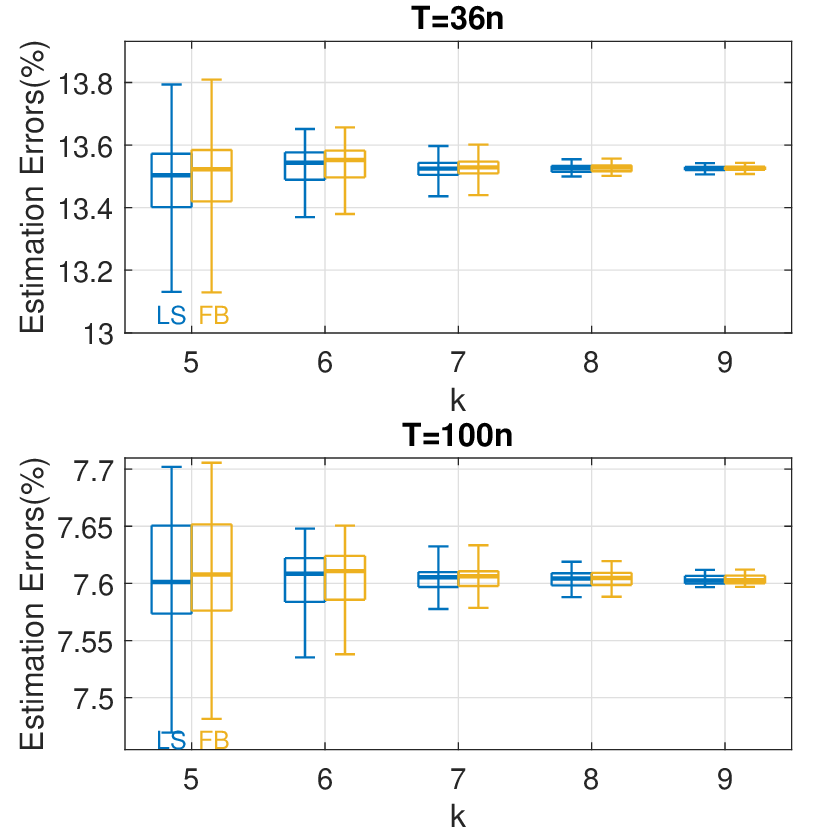}  
\caption{ {{Box plots of estimation errors $e_{RLS},e_{RFB}$ 
plotted against $k=\log_2(n/6)$: 
The errors of both estimators distribute very similarly and their medians 
converge to constants as the order $n$ grows with $T$ being a multiple of $n$.}} } \label{errFhigh}
\end{center}                               
\end{figure}

\begin{figure}
\begin{center}
\includegraphics[width=7.2cm]{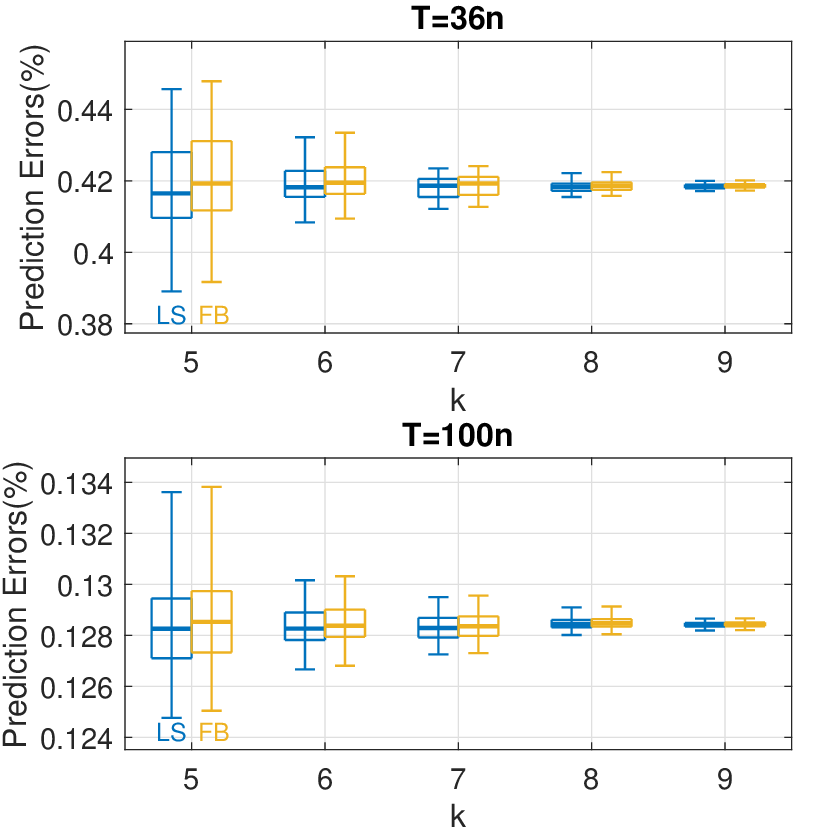}    
\caption{ {{Box plots of prediction errors $\epsilon_{RLS},\epsilon_{RFB}$ plotted against $k=\log_2(n/6)$: 
the observation is analogous to that of Fig. 6.}}}  
\label{errprehigh}
\end{center}                                 
\end{figure}

\section{Appendix: Proofs}
\setcounter{equation}{0}
We use the following reduced-rank least squares lemma.

{\bf Lemma R}. \cite{Rein98}[Theorem 2.2] 
\begin{align*}
\hat C&=\arg\min_{C_{n\times n}:\rank(C)=m}
\trace\{(Z-CX)' P^{-1} (Z-CX)\}\\
&=P^{\frac12} \hat V_m\hat V_m'P^{-\frac12} ZX'(XX')^{-1},
\end{align*}
where $\hat V_m=[\hat v_1,\dotsm,\hat v_m]$ with 
$\hat v_r$ the eigenvector corresponding to the $r$-th largest eigenvalue of 
the matrix
$\hat R=P^{-\frac12} ZX'(XX')^{-1} XZ'P^{-\frac12}$.

{\bf Proof of Theorem 4B. }\\ 
We show that $J(F; S_{11})$ has the form specified in Lemma R.
First, reorganize $J(F;P)$ to get 
 {$J(F;P) =  \trace\{P^{-1}(S_{w,f}+S_{w,b})\} = \trace\{P^{-1}(-4 S_{10} F' + F S_{00} F' + FPF'P^{-1} S_{11})\},$ }
where we dropped a constant term not dependent on $F$. 
Now, set $P= S_{11}$ to get 
\begin{align*}
	&J(F; S_{11}) \\
= 	&\trace\{ S_{11}^{-1}(-4 S_{10} F' + F( S_{00}+ S_{11})F'\}\nonumber\\
=	&\Vert  S_{11}^{-\frac12}(2 S_{10}( S_{00}+ S_{11})^{-\frac12} - F( S_{00}+ S_{11})^{\frac12})\Vert^2,
\end{align*}
where again we dropped a constant term not dependent on $F$. 
Now apply Lemma R to get the first quoted result.

We now prove that $Q_*= S_{11}^{-1}-\hat F_R' S_{11}^{-1}\hat F_R$ is positive definite.
Then Lyapunov's theorem gives that $\hat F_R$ is stable.
We have
\begin{align*}
Q_*&=
( S_{11}^{-1}-\hat F_{11}' S_{11}^{-1}\hat F_{11})+(\hat F_{11}' S_{11}^{-1}\hat F_{11}-\hat F_R' S_{11}^{-1}\hat F_R)\\
&=
 S_{11}^{-1} U_b S_{11}^{-1}+(\hat F_{11}' S_{11}^{-1}\hat F_{11}-\hat F_R' S_{11}^{-1}\hat F_R), 
\end{align*}
 {where 
$U_b=S_{11}-\hat F_b S_{11}\hat F_b'$, and 
$\hat F_b= S_{11}\hat F_{11} ' S_{11}^{-1}$.} 
For the second term we have
 {$
\hat F_{11} ' S_{11}^{-1}\hat F_{11} - \hat F_R' S_{11}^{-1}\hat F_R
=	\hat F_{11} ' S_{11}^{-\frac12}(I-\hat V_m\hat V_m') S_{11}^{-\frac12}\hat F_{11}\geq0.
$}
For the first term, we have
\begin{align*}\begin{array}{rrcl}
	&2 S_{10}&=&\hat F_{11}  S_{00} + \hat F_{11} S_{11}\\
\Rightarrow	&2 S_{01}&=& S_{00}\hat F_{11} ' +  S_{11}\hat F_{11} '\\
\Rightarrow 	&2 S_{01}&=& S_{00} S_{11}^{-1}\hat F_b  S_{11} + \hat F_b S_{11}\\
\Rightarrow 	&- S_{00} S_{11}^{-1}\hat F_b S_{11}\hat F_b ' &=& -2 S_{01}\hat F_b ' + \hat F_b  S_{11}\hat F_b '\\
\Rightarrow 	&-\Phi U_b &=&  S_{00} - 2 S_{01}\hat F_b + \hat F_b  S_{11}\hat F_b ' {{,}}
\end{array}\end{align*}
where $\Phi=- S_{00} S_{11}^{-1}$.
Now take the transpose and add the two equations together to get
 {$-\Phi U_b+ U_b(-\Phi)'
=
2[ S_{00}- S_{01}\hat F_b'-\hat F_b S_{10}+\hat F_b S_{11}\hat F_b']
=
\frac{2}{T}( Y_0-\hat F_b Y_1)( Y_0-\hat F_b Y_1)'=2\hat{S}_{w,b} {{,}} $}
which we show below is positive definite. 
But note that this is a continuous time Lyapunov equation
where $\Phi$ is a stability matrix - it has the 
same eigenvalues as $- S_{00}^{\frac12} S_{11}^{-1} S_{00}^{\frac12}$
which is negative definite with, therefore, negative real eigenvalues.
Thus $ U_b$ must be positive definite and the
result is established.

To see that $\hat{S}_{w,b}$ is positive definite first note that
from A1 and Theorem 2, $Q_b$ is positive definite.
Now  the backward least squares estimator
$\hat F_{b,LS}= S_{01} S_{11}^{-1}$ minimises $ S_{w,b}(F)$.
Then for any $F$, completion of squares gives
  {$S_{w,b}(F)=\hat Q_{b,LS}+(\hat F_{b,LS}-F) S_{11}(\hat F_{b,LS}-F)' {{,}}$} 
where $\hat Q_{b,LS}= S_{w,b}(\hat F_{b,LS})$. Set $F=0$ to get
$\hat Q_{b,LS}= S_{00}- S_{01} S_{11}^{-1} S_{10}$ which  {{converges in probability to }}$Q_b$.
Now since $ Q_b$ is full rank and the eigenvalues of $\hat Q_{b,LS}$ are
continuous functions of the entries in $\hat Q_{b,LS}$ they cannot
accumulate mass at $0$, i.e. the smallest eigenvalue of $\hat Q_{b,LS}$
is $0$ with zero probability. So $\hat Q_{b,LS}$ is positive definite w.p.1.
But setting $F=\hat F_b$ above, we see that
$\hat{S}_{w,b}\geq\hat  Q_{b,LS}$ and so is positive definite w.p.1.

{\bf Proof of Theorem 5E({{ii}})}.\\
We have 
\begin{align*}
\sqrt T(\hat F_{RLS}-\hat F_R)
=&
\sqrt T S_{11}^{\frac12}(\hat V_{*,m}\hat V_{*,m}' S_{11}^{-\frac12}\hat F_{LS}-\hat V_m\hat V_m' S_{11}^{-\frac12}\hat F_{11})\\
=&
 S_{11}^{\frac12}\hat V_{*,m}\hat V_{*,m}' S_{11}^{-\frac12}(\hat F_{LS}-\hat F_{11})\sqrt T\\
&+
 S_{11}^{\frac12}\sqrt T(\hat V_{*,m}\hat V_{*,m}'-\hat V_m\hat V_m') S_{11}^{-\frac12}\hat F_{11}.\end{align*}
From Theorem  {{5}}A, Lemma 5D, and the fact that $\Vert \hat V_{*,m}\hat V_{*,m}'\Vert^2=m,$
we get  {{that}} the first term  {{converges in probability to $0$}}.
For the second term, using Theorem 5A, Theorem 3B we are reduced to showing
$\sqrt T(\hat V_{*,m}\hat V_{*,m}'-\hat V_m\hat V_m')\xrightarrow{p} 0$.
Note that as in the proof of Theorem 5A we deduce that $\hat V_m\xrightarrow{p} V_m$.

Next set $\bar S=\frac12( S_{00}+ S_{11})$ and consider that
\begin{align*}
\begin{array}{rrcl}
&&&\sqrt T(\hat R-\hat R_*)\\
&&=&
\sqrt T S_{11}^{-\frac12}  S_{10}[\bar S^{-1}- S_{00}^{-1}] S_{10} ' S_{11}^{-\frac12}\\
&&=&
\sqrt T S_{11}^{-\frac12} S_{10} S_{00}^{-1}( S_{00}-\bar S)\bar S^{-1} S_{10}' S_{11}^{\frac12}\\
&&=&
\sqrt T S_{11}^{-\frac12} S_{10} S_{00}^{-1}\frac12( S_{00}- S_{11})\bar S^{-1} S_{10}' S_{11}^{-\frac12} {{,}}
\end{array}
\end{align*}
and we find $\sqrt T(\hat R-\hat R_*)\xrightarrow{p} 0$
in view of Theorem 5A and  {{L}}emma 5C.

Consider the `$k$-th' eigenvector $\hat v_k$ and eigenvalue $\hat \lambda_k$ of $\hat R$
as well as the corresponding $\hat v_{*,k}$ and $\hat \lambda_{*,k}$ of $\hat R_*$. 
Return now to $\sqrt T(\hat V_{*,m}\hat V_{*,m}'-\hat V_m\hat V_m')\xrightarrow{p} 0$. We rewrite this as
$\sum_{k=1}^{m}\sqrt T(\hat v_k\hat v_k'-\hat v_{*,k}\hat v_{*,k}')\xrightarrow{p} 0$ which holds if
$\sqrt T(\hat v_k\hat v_k'-\hat v_{*,k}\hat v_{*,k}')\xrightarrow{p} 0$ for each $k$. 
This has Frobenius norm $\sqrt T\sqrt{2[1-(\hat v_k'\hat v_{*,k})^2]}$.
If we set $\cos(\theta_k)=\hat v_k'\hat v_{*,k}$ then the norm is $\sqrt{2T}|\sin(\theta_k)|$.

This allows us to apply a special case of a classic result of \cite{DAVK70}
which is given in \cite{YU15}[the equation below their equation (1)] namely
 {$
|\sin(\theta_k)|\leq \frac1{\hat\delta_k}\Vert \hat R-\hat R_*\Vert,
$}
 {where $\hat\delta_k=\min(|\hat\lambda_{k-1}-\hat\lambda_{\ast,k}|,|\hat\lambda_{k+1}-\hat\lambda_{\ast,k}|)$. }
Since (Theorem 5A) $\hat\lambda_l\xrightarrow{p} \lambda_l,\hat\lambda_{\ast,l}\xrightarrow{p}\lambda_l$ for all $l$ then
$\hat\delta_k\xrightarrow{p} \delta_k=\min(|\lambda_{k-1}-\lambda_k|,|\lambda_{k+1}-\lambda_k|)$. 
However, consider that 
 {
$\sqrt T\Vert\hat R-\hat R_*\Vert=
\sqrt{\trace[\sqrt T(\hat R-\hat R_*)\sqrt T(\hat R-\hat R_*)]}, 
$}
but each term  {{converges in probability to $0$}} and so $\sqrt T|\sin(\theta_k)|\xrightarrow{p} 0$
 and the proof is complete.

\bibliographystyle{myplain}
\bibliography{srrvar1bib}  

\begin{thebibliography}{10}

\bibitem{Bert19}
D.~Bertsekas.
\newblock {\em Reinforcement Learning and Optimal Control}.
\newblock Athena Scientific, 2019.

\bibitem{Boot07}
B.~Boots, GJ. Gordon, \& S.~Siddiqi.
\newblock A constraint generation approach to learning stable linear dynamical
  systems.
\newblock {\em NIPS}, 20, 2007.

\bibitem{Burg75}
JP. Burg.
\newblock {\em Maximum Entropy Spectral Analysis}.
\newblock Stanford University, 1975.

\bibitem{RCHN20}
EY. Chen, RS. Tsay, \& R.~Chen.
\newblock Constrained factor models for high-dimensional matrix-variate time
  series.
\newblock {\em Jl. Am. Stat. Assocn.}, 115:775–793, 2020.

\bibitem{RCHN21}
R.~Chen, H.~Xiao, \& D.~Yang.
\newblock Autoregressive models for matrix-valued time series.
\newblock {\em Jl. Econom.}, 222:539–560, 2021.

\bibitem{Chui96}
NLC. Chui \& JM. Maciejowski.
\newblock Realization of stable models with subspace methods.
\newblock {\em Automatica}, 32(11):1587--1595, 1996.

\bibitem{DAVK70}
C.~Davis \& WM. Kahan.
\newblock The rotation of eigenvectors by a perturbation.
\newblock {\em SIAM Jl. Numer. Anal.}, 7:1--46, 1970.

\bibitem{Gill19}
N.~Gillis, M.~Karow, \& P.~Sharma.
\newblock {A}pproximating the nearest stable discrete-time system.
\newblock {\em Linear Algebra and its Applications}, 573:37--53,
  2019\color{black}.

\bibitem{Ham94}
JD. Hamilton.
\newblock {\em Time Series Analysis}.
\newblock Princeton Univ. Press, Princeton, NJ, 1994.

\bibitem{HORN13}
RA. Horn \& CA. Johnson.
\newblock {\em Matrix Analysis}.
\newblock Cambridge University Press, Cambridge, UK, 2013.

\bibitem{Jong23}
W.~Jongeneel, T.~Sutter, \& D.~Kuhn.
\newblock Efficient learning of a linear dynamical system with stability
  guarantees.
\newblock {\em IEEE Trans. on Autom. Contr.}, 68(5):2790--2804, 2023.

\bibitem{KAIL00}
T.~Kailath.
\newblock {\em Linear Estimation}.
\newblock Prentice Hall, Upper Saddle River, New Jersey, 2000.

\bibitem{Lutke05}
H.~Lutkepohl.
\newblock {\em Multivariate Time Series Analaysis}.
\newblock Springer, New York, 2005.

\bibitem{Neud99}
JR. Magnus \& H.~Neudecker.
\newblock {\em Matrix Differential Calculus}.
\newblock J. Wiley, New York, 1999.

\bibitem{Mall08}
G.~Mallet, G.~Gasso, \& S.~Canu.
\newblock New methods for the identification of a stable subspace model for
  dynamical systems.
\newblock In {\em IEEE Workshop on Machine Learning for Signal Processing},
  pages 432--437, 2008.

\bibitem{Mari00}
J.~Mari, P.~Stoica, \& T.~McKelvey.
\newblock Vector \uppercase{ARMA} estimation: a reliable subspace approach.
\newblock {\em IEEE Trans. Sig. Proc.}, 48(7):2092--2104, 2000.

\bibitem{Mill13}
DN. Miller \& RA. de~Callafon.
\newblock Subspace identification with eigenvalue constraints.
\newblock {\em Automatica}, 49(8):2468--2473, 2013.

\bibitem{Matn19}
M.~Nikolai, A.~Proutiere, A.~Rantzer, \& S.~Tu.
\newblock From self-tuning regulators to reinforcement learning and back again.
\newblock In {\em Proc. IEEE CDC}, pages 3724--3740, 2019.

\bibitem{Nofe21}
V.~Noferini \& F.~Poloni.
\newblock {N}earest {$\Omega$}-stable matrix via {R}iemannian optimization.
\newblock {\em Numerische Mathematik}, 148(4):817--851, 2021\color{black}.

\bibitem{Nutt77}
AH. Nuttall.
\newblock Multivariate linear predictive spectral analysis employing weighted
  forward and backward averaging: A generalization of \uppercase{B}urg's
  algorithm.
\newblock Report, Naval Underwater System Center, London, 1976.

\bibitem{Orba13}
FX. Orbandexivry, Y.~Nesterov, \& P.~{Van Dooren}.
\newblock {N}earest stable system using successive convex approximations.
\newblock {\em Automatica}, 49(5):1195--1203, 2013\color{black}.

\bibitem{Rech19}
B.~Recht.
\newblock A tour of reinforcement learning: The view from continuous control.
\newblock {\em Annual Review of Control, Robotics, and Autonomous Systems},
  2:253--279, 2019.

\bibitem{Rein98}
GC. Reinsel \& RP. Velu.
\newblock {\em Multivariate Reduced-Rank Regression: Theory and Applications},
  volume 136.
\newblock Springer, 1998.

\bibitem{Rong23}
X.~Rong \& V.~Solo.
\newblock State space subspace noise modeling with guaranteed stability.
\newblock In {\em Proc. IEEE CDC}, pages 4203--8208, 2023.

\bibitem{Stra77}
O.~Strand.
\newblock Multichannel complex maximum entropy (autoregressive) spectral
  analysis.
\newblock {\em IEEE Trans. Autom. Contr.}, 22(4):634--640, 1977.

\bibitem{Sutt18}
RS. Sutton \& AG. Barto.
\newblock {\em Reinforcement Learning: An Introduction}.
\newblock MIT press, 2018.

\bibitem{Tana05}
H.~Tanaka \& T.~Katayama.
\newblock Stochastic subspace identification guaranteeing stability and minimum
  phase.
\newblock {\em IFAC Proceedings Volumes}, 38(1):910--915, 2005.

\bibitem{Umen18}
J.~Umenberger, J.~Wågberg, IR. Manchester, \& TB. Schön.
\newblock Maximum likelihood identification of stable linear dynamical systems.
\newblock {\em Automatica}, 96:280--292, 2018.

\bibitem{VanG01}
T.~Van~Gestel, JAK. Suykens, P.~Van~Dooren, \& B.~De~Moor.
\newblock Identification of stable models in subspace identification by using
  regularization.
\newblock {\em IEEE Trans. Autom. Contr.}, 46:1416--1420, 2001.

\bibitem{VanO96}
P.~Van~Overschee \& B.~De~Moor.
\newblock {\em Subspace Identification for Linear Systems: Theory,
  Implementation, Applications}.
\newblock Kluwer, Boston, 1996.

\bibitem{Weik16}
C.~Weikert \& MB. Schulze.
\newblock Evaluating dietary patterns: the role of reduced rank regression.
\newblock {\em Current Opinion in Clinical Nutrition and Metabolic Care},
  19(5):341--346, 2016.

\bibitem{YU15}
Y.~Yu, T.~Wang, \& RJ. Samworth.
\newblock A useful variant of the {D}avis–{K}ahan theorem for statisticians.
\newblock {\em Biometrika}, 102:315–323, 2015.

\end{thebibliography}
\end{document}